\documentclass[12pt,a4paper]{article}
\usepackage{amsfonts,latexsym}
\usepackage{graphicx,color}
\usepackage{dcolumn}
\usepackage{graphicx}
\usepackage{amsmath}
\usepackage{amsfonts}
\usepackage{amssymb}

\renewcommand{\thefootnote}{\fnsymbol{footnote}}

\begin{document}

\vspace{12mm}

\begin{center}
{{{\Large {\bf   Scalarized charged  black holes with scalar mass term }}}}\\[10mm]

{ De-Cheng Zou$^{a,b}$\footnote{e-mail address: dczou@yzu.edu.cn} and Yun Soo Myung$^a$\footnote{e-mail address: ysmyung@inje.ac.kr} }\\[8mm]

{${}^a$Institute of Basic Sciences and Department  of Computer Simulation, Inje University Gimhae 50834, Korea\\[0pt] }

{${}^b$Center for Gravitation and Cosmology and College of Physical Science and Technology, Yangzhou University, Yangzhou 225009, China\\[0pt]}
\end{center}
\vspace{2mm}

\begin{abstract}
We  study the scalarized charged black holes  in the Einstein-Maxwell-Scalar (EMS) theory with scalar mass term.
In this work, the scalar mass term is chosen to be $m^2_\phi=\alpha/\beta$, where  $\alpha$ is a coupling parameter and $\beta$  is a mass-like parameter.
It turns out that any scalarized charged black holes are not allowed for the case of  $\beta \le 4.4$ with $q=0.7(M=0.5,Q=0.35)$ because this case implies   the stable Reissner-Nodstr\"{o}m (RN) black holes. In the massless limit of $\beta\to \infty$, one recovers the case of the EMS theory.
We note that the unstable RN black hole implies the appearance of scalarized charged black holes. The other unstable case of $\beta>4.4$ with $q=0.7$ allows us to obtain the $n=0,1,2,\cdots$ scalarized charged black holes for  $\alpha(\beta)\ge \alpha_{\rm th}(\beta)$ where  $\alpha_{\rm th}(\beta)$ represents the threshold of instability for the RN black hole. Furthermore, it is shown that
the $n=0$ black hole is stable against radial perturbations, while the $n=1$ black hole is unstable. This stability result is independent of the mass parameter $\beta$.

\end{abstract}
\vspace{5mm}

\vspace{1.5cm}

\hspace{11.5cm}
\newpage
\renewcommand{\thefootnote}{\arabic{footnote}}
\setcounter{footnote}{0}

%%%% Introduction %%%%

\section{Introduction}
Recently, the scalarized black holes have been found from the Einstein-Gauss-Bonnet-Scalar (EGBS) theory which includes
a scalar-Gauss-Bonnet coupling term [$f(\phi){\cal G}$]~\cite{Doneva:2017bvd,Silva:2017uqg,Antoniou:2017acq}.
Here, $f(\phi)=\alpha \phi^2/2$ is chosen  for quadratic coupling and $f(\phi)= \alpha(1-e^{-6\phi^2})/12$ for exponential coupling, indicating different types from linear and  dilatonic couplings.
In these models, the appearance of scalar hairy black hole  results from the unstable Schwarzschild black holes.
The  scalar-Gauss-Bonnet coupling term induces negative potential  outside the horizon  and the coupling constant $\alpha$ plays the role of a spectral parameter
in the linearized scalar equation around the Schwarzschild black hole.
It is meaningful to say that the scalar hairy black holes appear through a spontaneous scalarization for a large  coupling constant in the full EGBS system.

More recently, introducing a scalar mass term  has an effect  on the bifurcation points
where the scalarized black holes branch out of the Schwarzschild black hole without scalar hair in the EGBS theory~\cite{Brihaye:2018grv,Macedo:2019sem,Doneva:2019vuh}. This theory includes a quadratic scalar term with  mass as well as the scalar-Gauss-Bonnet coupling term. In other words, the mass term changes the threshold for scalarization surely and  it may give the black hole  mass range over which scalarized black holes can exist.  Moreover, it is suggested  that a quartic scalar term is sufficient to make a stable $n=0$ black hole against the radial  perturbations without introducing  an exponential coupling term. However, we note that this indicates  a feature of the EGBS theory with quadratic coupling.
In this direction, it is worth noting that the scalarized charged black holes were found from the  Einstein-Maxwell-Scalar (EMS) theory~\cite{Herdeiro:2018wub,Fernandes:2019rez}.
We would like to mention  that  the $n=0$ black hole are stable in the EMS theory with exponential and quadratic couplings~\cite{Myung:2019oua}.

On the other hand, it is  curious to know why a single branch of the non-Schwarzschild black hole with Ricci tensor hair exists in the Einstein-Weyl (EW) gravity whose Lagrangian takes the form of ${\cal L}_{\rm EW}=\sqrt{-g}[R-C_{\mu\nu\rho\sigma}C^{\mu\nu\rho\sigma}/2m^2_2]$ with coupling parameter $m^2_2$~\cite{Lu:2015cqa}. Actually, an apparent difference implies that many branches of $\alpha_{n=0,1,2,\cdots}= \{8.019, 40.84, 99.89,\cdots\}$ for
$q = 0.7$ exist in the EMS theory with exponential coupling term ($e^{\alpha \phi^2}F^2$), while  a single branch of $m^2_2
= 0.767$ exists for the EW gravity~\cite{Lu:2017kzi}. This seems to appear  because an asymptotic form of Zerilli potential ($V_{\rm Z}\to m^2_2$) is different from the scalar potential ($V\to0$) in the EMS theory~\cite{Myung:2018vug}. This means  that  the scalar perturbation  vanishes   asymptotically ($\phi_\infty\to 0$) in the EMS theory, while the $s(l = 0)$-mode of Ricci tensor perturbation takes
a normalizable form ($\psi_{\infty}\to e^{-m_2 r}$) in the EW gravity.  An asymptotic  correspondence would be met naively when one proposes a mass term of
$V_\phi=2m^2\phi^2$. In this case,  a scalar potential takes the form
of $V_{\rm mass}(r) = f(r)[2M/r^3+m^2-(m^2+2)Q^2/r^4]$
which shows a similar asymptote ($V_{\rm mass}\to m^2$, as $r\to \infty$) to $V_{\rm Z}(r)$. However, it turns out that for this mass term,  all  potentials are positive definite outside the horizon, providing the sufficient condition for stability.  Therefore, this choice does not allow any scalarized charged black holes.  Of course, an independent choice of mass parameter is available  and it may lead to the scalarized charged black holes  on the analogy of the EGBS theory with scalar mass term.

In this work, we wish to investigate how  the number of bifurcation points can be changed when including a specific mass term of $V_\phi=2(\alpha/\beta)\phi^2$. Here, $\alpha$ is a coupling parameter and $\beta$ is a mass parameter in the EMS theory with scalar mass term. The original motivation is mainly to explain  a difference between many branches in the EMS theory and  a single branch in the EW theory.
However, it turns out that for $\beta>4.4$ with $q=0.7$,
the number of bifurcation points  remains unchanged when  including  such a mass term.
Instead, for $\beta\le4.4$ with the same $q$, there is no unstable RN black hole and thus, one could not find any scalarized charged black holes. In the massless limit of $\beta\to \infty$, one recovers the case of the EMS theory. This implies that the role of scalar mass term provides either nothing or all bifurcation points, but it does not  lead  to a single branch of scalarized charged black holes. This indicates  a difference between scalar and tensor hairs.
Finally, we show that
the $n=0$ black hole is stable against radial perturbations, while the $n=1$ black hole is unstable.  This result is independent of the mass parameter $\beta$.

\section{EMS theory}

The EMS theory with scalar mass term takes the form ~\cite{Herdeiro:2018wub}
\begin{equation}
S=\frac{1}{16 \pi}\int d^4 x\sqrt{-g}\Big[ R-2\nabla_\mu \phi \nabla^\mu \phi-2m^2_\phi \phi^2-e^{\alpha \phi^2} F^2\Big],\label{Action1}
\end{equation}
where the mass squared   is chosen to be $m^2_\phi=\alpha/\beta>0$. Here $\alpha$($\beta$) are  coupling (mass) parameters and the type of scalar coupling to the Maxwell term is exponential. The other case of $m^2_\phi<0(\beta<0)$ corresponds to a genuinely tachyonic instability and, therefore, this will be excluded from our consideration.
First, we derive  the Einstein  equation
\begin{eqnarray}
 G_{\mu\nu}=2\nabla _\mu \phi\nabla _\nu \phi -\Big[(\nabla \phi)^2+\frac{\alpha}{\beta}\phi^2\Big]g_{\mu\nu}+2e^{\alpha \phi^2}T_{\mu\nu} \label{equa1}
\end{eqnarray}
with $G_{\mu\nu}$  the Einstein tensor and  $T_{\mu\nu}=F_{\mu\rho}F_{\nu}~^\rho-F^2g_{\mu\nu}/4$.
The Maxwell equation is coupled to scalar as
\begin{equation} \label{M-eq}
\nabla^\mu F_{\mu\nu}-2\alpha \phi\nabla^{\mu} (\phi)F_{\mu\nu}=0.
\end{equation}
We obtain  the scalar field  equation
\begin{equation}
\nabla^2 \phi -\frac{\alpha}{\beta}\phi-\frac{\alpha}{2} e^{\alpha \phi^2}F^2 \phi=0 \label{s-equa}.
\end{equation}

Taking into account   $\bar{\phi}=0$ and electrically charged  $\bar{A}_t=Q/r$,  the RN  solution is found  when solving  (\ref{equa1}) and (\ref{M-eq})
\begin{equation} \label{ansatz}
ds^2_{\rm RN}= \bar{g}_{\mu\nu}dx^\mu dx^\nu=-f(r)dt^2+\frac{dr^2}{f(r)}+r^2d\Omega^2_2
\end{equation}
with the metric function
\begin{equation}
f(r)=1-\frac{2M}{r}+\frac{Q^2}{r^2}.
\end{equation}
The outer (inner) horizon is located at $r=r_\pm=M(1\pm\sqrt{1-q^2})$ with $q=Q/M$.
We stress that the RN solution (\ref{ansatz}) is a black hole solution to the EMS theory with scalar mass term, being independent of $\alpha$ and $\beta$.
Hereafter, we will choose a particular  case of $q=0.7(M=0.5,Q=0.35)$ as a representative of non-extremal RN black holes.
In this case, solving $f(r)=0$ determines the outer horizon at $r=r_+=0.857$ and the inner horizon $r=r_-=0.143$.

Finally, we would like to note that the case of $\bar{\phi}$=const may provide a different solution because their equations are given by
\begin{equation}
\bar{G}_{\mu\nu}=-\frac{\alpha}{\beta}\bar{\phi}^2g_{\mu\nu}+2e^{\alpha \bar{\phi}^2}\bar{T}_{\mu\nu},~~\bar{\nabla}^\mu \bar{F}_{\mu\nu}=0,~~\frac{1}{\beta}=-\frac{1}{2} e^{\alpha \bar{\phi}^2}\bar{F}^2.
\end{equation}
In this case, the last relation reduces to
\begin{equation}
\frac{1}{\beta}=e^{\alpha \bar{\phi}^2}\frac{Q^2}{r^4}
 \end{equation}
 which means that $\beta$ is not a proper coupling constant. So, we exclude the case of $\bar{\phi}$=const from our consideration.

\section{Stability for  RN black hole }
The linearized theory around the RN black hole could be obtained to investigate the stability analysis of a  RN black hole with $q=0.7$.
The perturbed fields are introduced by considering metric tensor ($g_{\mu\nu}=\bar{g}_{\mu\nu}+h_{\mu\nu}$), vector ($A_\mu=\bar{A}_\mu+a_\mu$), and scalar ($\phi=\bar{\phi}+\varphi$) with $\bar{\phi}=0$. We note that  there  are two ways to obtain the linearized theory. One way is first to bilinearize the action (\ref{Action1}) and then, obtain its linearized equations by varying perturbed fields. The other is to linearize equations (\ref{equa1})-(\ref{s-equa}) directly.  Adapting  the latter leads to the linearized Einstein-Maxwell  equations
\begin{eqnarray}
 \delta G_{\mu\nu}(h) = 2\delta T_{\mu\nu},~~
  \bar{\nabla}^\mu f_{\mu\nu}=0 \label{l-eq1}
\end{eqnarray}
with a decoupled scalar  equation
\begin{equation}
\Big[\bar{\nabla}^2-\frac{\alpha}{\beta}+ \alpha \frac{Q^2}{r^4}\Big] \varphi= 0. \label{l-eq2}
\end{equation}
In the EMS theory, the last term in (\ref{l-eq2}) develops  a negative potential outside the horizon and thus, it may induce the instability.
In the EMS theory with scalar mass term, however, there exists a competition between mass term and the last term to give a negative potential outside the horizon. Therefore, the instability is harder to realize for large scalar masses.
Concerning  the stability analysis  of the RN black hole,
we  consider  the two linearized equations in (\ref{l-eq1}) first  because two  of metric $h_{\mu\nu}$ and vector $a_{\mu}$   are coupled to each other  as in the Einstein-Maxwell theory.
It is worth noting that these are exactly the same linearized equations for the Einstein-Maxwell theory.
 We briefly review the stability of RN black hole in the Einstein-Maxwell theory.
In this case, one obtained  the Zerilli-Moncrief equation describing  two physical degrees of freedom (DOF)
for  the odd-parity perturbations~\cite{Zerilli:1974ai,Moncrief:1974gw}, while
the even-parity perturbations for two physical DOF were investigated in~\cite{Moncrief:1974ng,Moncrief:1975sb}.
It is known that the RN black hole is  stable against the tensor-vector perturbations.

Hence, the instability of RN black holes in  the EMS theory with scalar mass term will be determined entirely by the
linearized scalar equation (\ref{l-eq2}),  indicating a feature of the EMS theory with scalar mass term.
Now, let us introduce   the separation of variables around a spherically symmetric RN background (\ref{ansatz})
\begin{equation} \label{scalar-sp}
\varphi(t,r,\theta,\chi)=\frac{u(r)}{r}e^{-i\omega t}Y_{lm}(\theta,\chi).
\end{equation}
Choosing a tortoise coordinate $r_*$ defined by $r_*=\int dr/f(r)$, a radial part of the scalar equation takes the form
\begin{equation} \label{sch-2}
\frac{d^2u}{dr_*^2}+\Big[\omega^2-V(r)\Big]u(r)=0.
\end{equation}
Here the scalar potential $V(r)$ is given by
\begin{equation} \label{pot-c}
V(r)=f(r)\Big[\frac{2M}{r^3}+\frac{l(l+1)}{r^2}+\frac{\alpha}{\beta}-\frac{2Q^2}{r^4}-\alpha\frac{Q^2}{r^4}\Big],
\end{equation}
which seems to be a complicated form.
The $s(l=0)$-mode is an allowable mode  for the scalar perturbation and thus,  it could be  used to test the instability of the RN black hole.
Hereafter, we confine ourselves  to the $l=0$ mode. It is interesting to note that $V(r) \to \alpha/\beta$ as $r\to \infty$, compared   to the massless potential of  $V_{\beta\to \infty}(r)\to 0$ in the EMS theory.
 From the potential (\ref{pot-c}),  the condition for  positive definite potential which corresponds to sufficient condition for stability  could be found  as~\cite{Myung:2019oua}
\begin{equation}
V(r) \ge 0 \to \beta \le G(r,\alpha)= \frac{\alpha r^4}{Q^2(\alpha+2)-2Mr}.\label{c-po}
\end{equation}
%%%%%%%%%%%%%%%%%%%%%%%%%%%%%%%%%%%%%%%%%%%%%%%%%%%%%%%%%%%%%%%%%%%%%%%%%%%%%%
We observe the behavior  of $G(r,\alpha)$ function with $M=0.5$ and $Q=0.35$  pictorially.
Its minimum stays near $r=r_+$ as $\alpha$ increases for $r\in[r_+=0.857,2]$  and $\alpha \in [0.01,100]$.  A minimum value of $G(r,\alpha)$ locates at 5 around $r=r_+$ for $\alpha=1000$.
We read off the stability bound  from $G(r,\alpha)$ as
\begin{equation} \label{st-con}
\beta \le G(r_+,\alpha\to \infty)=\frac{r_+^4}{Q^2}=4.4.
\end{equation}
However, it is not easy to obtain the instability condition  from the potential (\ref{pot-c}) directly.
In this direction, we need to look for  the negative region of potential outside the horizon because it may indicate a signal of instability.
Taking into account  the stability condition (\ref{st-con}), one expects that a negative region may allow for $\beta>4.4$ and $\alpha<\infty$.
As an  example, we wish to display the negative region of potential (\ref{pot-c}) as function of $r$ and $\alpha$ for $\beta=811$ in Fig. 1(Left).  We find from Fig. 1(Right) that the width and depth of negative region in $V(r,\alpha)$ increase as $\alpha$ increases.
%%%%%%%%%%%%%%%%%%%%%%%%%%%%%%
\begin{figure*}[t!]
   \centering
  \includegraphics{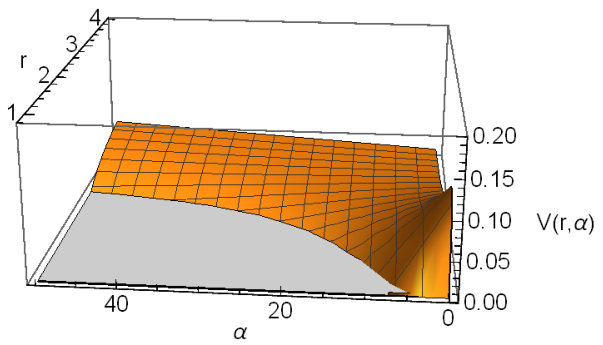}
     \hfill%
   \includegraphics{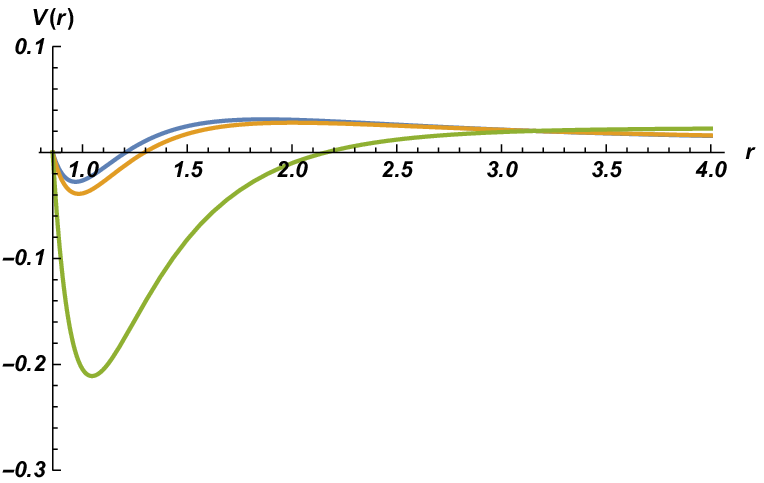}
\caption{(Left)The $\alpha$-dependent potential $V(r,\alpha,\beta=811)$ as function of $r \in [r_+,4.0]$ and $\alpha \in[0.01,50]$ for $q=0.7$. The shaded region
 along $\alpha$-axis represents negative region of the potential. (Right)Plots of three potentials $ V(r,\alpha,\beta=811)$  with  three different $\alpha=\{8,\alpha_{\rm th}=8.82,20\}$ from top to bottom near the $V$-axis. }
\end{figure*}
%%%%%%%%%%%%%%%%%%%%%%%%%%%%%%%%%%%%%%%%%%%%%%%%%%%%%%%%%%%%%%%%%%%%%%%%%%%%%%
It is conjectured that  if the potential $V(r)$  is negative in some region, a growing perturbation may appear in the spectrum, indicating  an instability of a RN black hole.
However, this is not always true. A determining condition for whether a black hole
is stable or not   depends on whether the time-evolution of the scalar perturbation is decaying or not.
The linearized  scalar equation (\ref{sch-2}) around RN black hole may allow  an unstable (growing) mode like $e^{\Omega t}$ for a scalar perturbation and thus, it  indicates the sign for instability of the black hole.  Importantly, it is stated  that the instability of RN black holes implies the appearance of scalarized charged black holes.
Therefore, we have to solve (\ref{sch-2}) after replacing $\omega =-i\Omega$ numerically by imposing   boundary conditions: purely ingoing wave near the horizon and  purely outgoing wave at infinity.
From Fig. 2, we read off the threshold of instability [$\alpha_{\rm th}(\beta)$].
Hence, the instability bound can be determined   numerically by
\begin{equation}
\alpha(\beta) \ge \alpha_{\rm th}(\beta) \label{In-cond}
\end{equation}
with $\alpha_{\rm th}(\beta)=\{9.345(356),~8.82(811),~8.60(1400),~8.019(\infty)\}$.  On the other hand, one always finds stable RN black holes for $\alpha(\beta)<\alpha_{\rm th}(\beta)$.
%%%%%%%%%%%%%%%%%%%%%%%%%%%
\begin{figure*}[t!]
   \centering
   \includegraphics{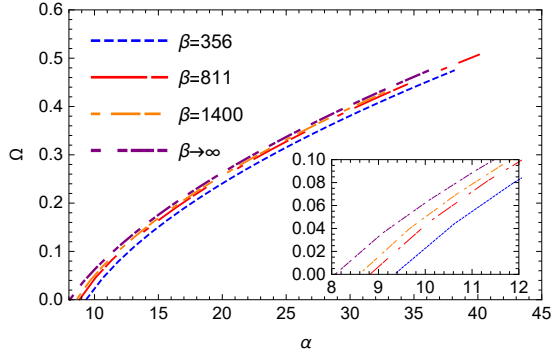}
\caption{Four graphs of $\Omega$ in $e^{\Omega t}$ as functions of  $\alpha$ are used to determine the thresholds of instability [$\alpha_{\rm th}(\beta)$] around RN black hole. These correspond to the crossing points at $\alpha$-axis appearing in the magnification of  the enclosed region. We find $\alpha_{\rm th}(\beta)=9.345(356),~8.82(811),~8.60(1400),~8.019(\infty)$. The last one denotes the massless limit (the EMS theory) at $\beta=\infty$, whereas the other stable boundary at  $\beta=4.4$ could not represent here because it appears at $\alpha=\infty$.}
\end{figure*}
From (Right) Fig. 1, for $\beta=811$, one finds  $\alpha<\alpha_{\rm th}=8.82$ for stable RN black hole  and $\alpha\ge \alpha_{\rm th}$ for unstable RN black holes.

\section{Static scalar  perturbation: bifurcation points}
\begin{table}[h]
\resizebox{14cm}{!}
{\begin{tabular}{|c|c|c|c|c|c|c|c|c|c|c|c|c|c|c|c|c|c|c|}
  \hline
$\beta$&4.4 &$\cdots$ & 5.5 & 6 &8 &20 &33 &71 & 125 &\underline{258} & \underline{356} & 471& 591 &\underline{ 811} &\underline{1400}&\underline{2000}&$\cdots$ &$\infty$\\ \hline
$\alpha_{n=0}(\beta)$&$\infty$ &$\cdots$&469.8 & 256.6 & 82.52  &21.80 &15.98 & 12.10 &10.69 & 9.649 & 9.345& 9.132 &8.987&8.82 &8.60&8.493&$\cdots$&8.019\\ \hline
$\alpha_{n=1}(\beta)$&$\infty$  &$\cdots$&4074& 2194 & 675.5 & 158.8& 77.69 &110.1 &65.77 & 56.73 &54.01 & 52.07 & 50.73& 49.15 &47.03&45.96&$\cdots$&40.84 \\
\hline
\end{tabular}}
\caption{List for the first two bifurcation points depending on $\beta$: $\alpha_{n=0}(\beta)$ represents the fundamental branch and $\alpha_{n=1}(\beta)$ denotes the first excited branch. In the limit of $\beta \to 4.4$, one recovers  stable RN black hole, while  one recovers  $\alpha_{n=0,1}$ for the EMS theory in the massless limit of $\beta \to  \infty$. The underlined cases are used for stability analysis. }
\end{table}
\begin{figure*}[t!]
   \centering
   \includegraphics{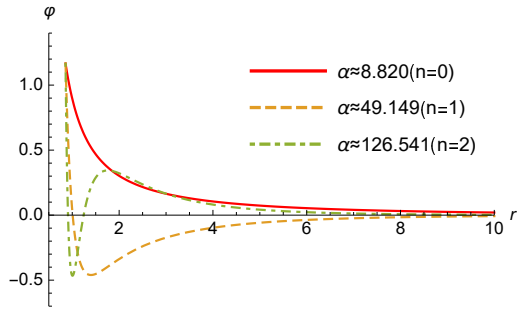}
\caption{Plot of radial profiles  $\varphi(z)=u(z)/z$ as function of $z=r/2M$ for the first three static perturbed scalar solutions with $\beta=811$ and $q=0.7$.
The number $n$  of zero nodes describes the $n=0,1,2$ scalarized charged black holes. }
\end{figure*}
%%%%%%%%%%%%%%%%%%
Now, let us  check the instability bound (\ref{In-cond}) again because  the precise value of  $\alpha_{\rm th}(\beta)$ determines the appearance of  scalarized charged  black holes.
This can be confirmed by obtaining  a static scalar solution [scalar cloud: $\varphi(r)$] to the linearized equation (\ref{sch-2}) with $u(r)=r\varphi(r)$ and $\omega=0$ on the RN  background.
For a given $l=0$ and $q=0.7$, requiring an asymptotically normalizable solution ($\varphi_\infty \to e^{-\sqrt{\alpha/\beta}r}/r$) leads to the fact that  the existence  of a smooth scalar  determines  a discrete set for $\alpha_{n}(\beta)$ where $n=0,1,2,\cdots$ denotes the number of zero crossings for $\varphi(r)$ (or order number). See Fig. 3 for static scalar solutions $\varphi(z)$ for $z=r/2M$ with $\beta=811$ and $q=0.7$.
The $n=0$ scalar mode  represents the fundamental branch of scalarized charged  black holes,
while the $n=1,2$ scalar modes  denote $n=1,2$ higher branch of scalarized charged black holes.
It is noted that  this corresponds to finding the first three bifurcation points from the RN black hole.

\begin{figure*}[t!]
   \centering
   \includegraphics{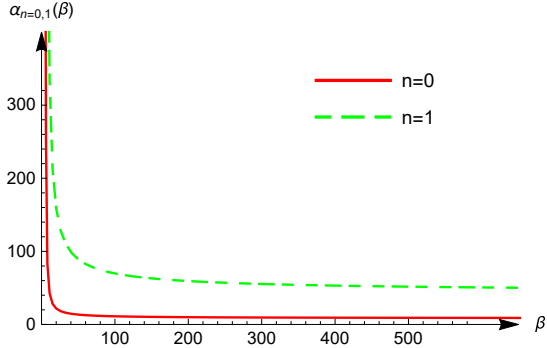}
\caption{Plot of  $\alpha_{n=0,1}(\beta)$ based on Table 1 shows effects of the mass term $\beta$ on the scalarization. Observing  $\alpha_{n=0}(\beta\to 4.4)\to \infty$ implies  that unstable RN black holes exit for $\beta> 4.4$. This implies  that any scalarized charged black holes would  be found for $\beta> 4.4$. Also, we note that  $\alpha_{n=0,1}(\beta\to \infty)$ reduces to   $\alpha_{n=0,1}=\{8.019,40.84\}$  for the EMS theory.  }
\end{figure*}
Consequently, we confirm from Fig. 2 and Table 1 that for given $\beta$,
\begin{equation}
\alpha_{\rm th}(\beta)=\alpha_{n=0}(\beta)
\end{equation}
which states that the threshold of instability for RN black hole is precisely the appearance of $n=0$ scalarized charged black holes.
We find from Fig. 4 that $\alpha_{n=0,1}(\beta)$ increases as $\beta$ decreases. The instability is therefore harder to realize for larger scalar masses (as $\beta \to 4.4$).
This picture is similar to Fig. 1(Left) in Ref.~\cite{Macedo:2019sem}, where no upper limit appears because they used an independent mass term.
 We find that in the massless limit of $\beta\to\infty$, $\alpha_{n=0,1}(\beta)$ approaches $\alpha_{n=0,1}=\{8.019,40.84\}$ for the EMS theory.
 Also, we observe  the other limit  that $\alpha_{n=0}(\beta)\to \infty$, as $\beta \to 4.4$.
In other words, we show that unstable RN black holes exist for $\beta> 4.4$ [see the opposite bound (\ref{st-con}) for stable RN black holes].
This implies that the $n=0$  scalarized charged black hole would  be found for $\alpha\ge\alpha_{n=0}(\beta)$ with $8.019\le \alpha_{n=0}(\beta) <\infty $ for $\beta\in (4.4,\infty]$, showing a significant shift of the $n=0$ scalarized charged black holes in compared to the massless case ($\alpha\ge \alpha_{n=0}(\beta \to\infty)=8.019$)  in the EMS theory.
Particularly, an unallowable region for scalarization is given by $0<\beta\le 4.4$ where the unstable RN black holes are never found for any $\alpha>0$.  Finally, we note that the case of $m^2_\phi=2\alpha$ with $\beta=1$ corresponds to the stable RN black hole. Therefore, one could not find any scalarized black holes from this case.

\section{Scalarized charged black holes}

First of all, we would like to mention   that the RN black hole  is allowed for any value of $\alpha$,
while a scalarized charged  black hole solution
may exist only for  $\alpha(\beta)\ge \alpha_{\rm th}(\beta)$ and $\beta >4.4$.
The threshold of  instability for a RN black hole  denotes an exact appearance of the $n=0$ scalarized charged black hole.
So, we derive the $n=0$  scalarized RN black hole for $q=0.7$ and $\alpha(\beta=811)=8.82 \ge \alpha_{n=0}(\beta=811)=8.82$ case numerically.
For this purpose, let us introduce  a spherically symmetric metric ansatz as
\begin{eqnarray}\label{nansatz}
ds^2_{\rm SCBH}=-A(r)dt^2+\frac{dr^2}{B(r)}+r^2(d\theta^2+\sin^2\theta d\chi^2).
\end{eqnarray}
Also, we consider the $U(1)$ potential $A_\mu=\{v(r),0,0,0\}$ and  scalar $\phi(r)$.
Substituting these  into Eqs.(\ref{equa1})-(\ref{s-equa}) leads to   four equations for $\{A(r),B(r),v(r),\phi(r)\}$ as
\begin{eqnarray}
&&\frac{1}{r^2}+\frac{1}{B}(-\frac{1}{r^2}+\frac{\alpha \phi^2}{\beta})+\frac{A'+e^{\alpha \phi^2}r v'^2}{r A}-\phi'^2=0,\label{neom1}\\
&&-\frac{\alpha  \phi^2}{\beta}+\frac{1-B-rB'}{r^2}-e^{\alpha \phi^2} \frac{ Bv'^2}{A}-B \phi'^2=0,\label{neom2}\\
&&Q+e^{\alpha \phi^2}r^2\sqrt{\frac{B}{A}}v'=0,\label{neom3}\\
&& \phi''+\Big(\frac{2}{r}+\frac{A'}{2A}+\frac{B}{2 B'}\Big)\phi'+\Big(-\frac{\alpha}{\beta B}+\frac{\alpha e^{\alpha \phi^2} v'^2}{A}\Big)\phi=0. \label{neom4}
\end{eqnarray}
One finds an
approximate solution to equations in the near horizon
\begin{eqnarray}
&&A(r)=A_1(r-r_+)+A_2(r-r_+)^2+\ldots,\label{aps-1}\\
&&B(r)=B_1(r-r_+)+B_2(r-r_+)^2+\ldots,\label{aps-2}\\
&&\phi(r)=\phi_0+\phi_1(r-r_+)+\ldots,\label{aps-3}\\
&&v(r)=v_1(r-r_+)+v_2(r-r_+)^2+\ldots\label{aps-4}
\end{eqnarray}
with the first-order three coefficients
\begin{eqnarray}
&&B_1=\frac{1}{r_+}\Big(1-\frac{Q^2e^{-\alpha \phi^2_0}}{r_+^2}-\frac{\alpha r_+^2 \phi_0^2}{\beta}\Big),~~
\phi_1=\frac{\alpha(Q^2\beta-r_+^4e^{\alpha \phi^2_0})\phi_0}{Q^2r_+\beta+r_+^3(-\beta+\alpha r_+^2\phi^2_0)e^{\alpha \phi^2_0}}, \label{ncoef} \\
&&\quad v_1=-\frac{e^{-\alpha\phi_0^2}Q\sqrt{A_1}}{\sqrt{r_+(r_+^2-e^{-\alpha \phi^2_0}Q^2-\frac{\alpha r_+^4 \phi_0^2}{\beta})}}. \nonumber
\end{eqnarray}
Here $A_1$ is a free parameter.   $\phi_0=\phi(r_+)$  will be
determined  when  matching  (\ref{aps-1})-(\ref{aps-4}) with the asymptotic  solutions in the far region of $r \gg r_+$
\begin{eqnarray}
&&A(r \gg r_+)=1-\frac{2M}{r}+\ldots,\quad  B(r\gg r_+)=1-\frac{2M}{r}+\ldots, \nonumber \\
&& \phi(r \gg r_+)=\phi_{\rm ml}e^{-\sqrt{\frac{\alpha}{\beta}} r}+\ldots,\quad  v(r \gg r_+)=\Phi+\frac{Q}{r}+\ldots,\label{asym-sol}
\end{eqnarray}
where $\phi_{\rm ml}=Q_s/r$ denotes the scalar hair for the EMS theory and $\Phi=Q/r_+$ denotes the electrostatic potential. In addition, $M,~Q_s,$ and $Q$ denote the ADM mass, the scalar charge, and the electric charge, respectively. In the massless limit of $\beta \to \infty$, one recovers the asymptotic solution for the EMS theory.
\begin{figure*}[t!]
   \centering
   \includegraphics{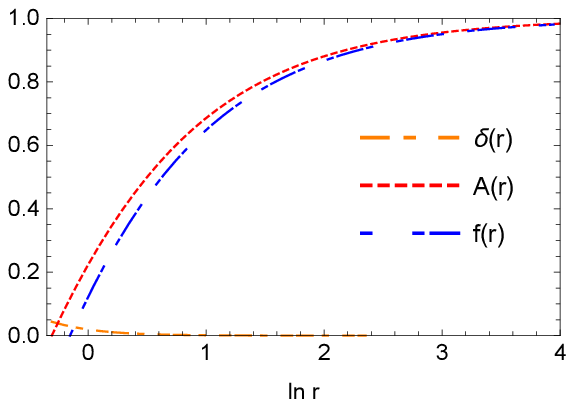}
   \hfill%
   \includegraphics{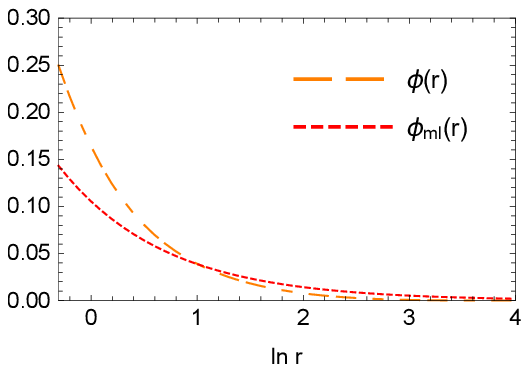}
\caption{Plots of a scalarized charged black hole with $\alpha=8.82$ in the $n=0$  branch of $\alpha(\beta=811) \ge 8.82$ and $q=0.7$. (Left) Metric function $\delta(r)=\ln[B(r)/A(r)]/2,~A(r)$, and $f(r)$ for the RN black hole.  (Right) Scalar hair $\phi(r)$ and  scalar hair  $\phi_{\rm ml}$ for the EMS theory with  scalar charge $Q_s=0.105$. }
\end{figure*}
%%%%%%%%%%%%%%%%%%

Consequently, we obtain the $n=0$ scalarized charged black hole solution shown in Fig. 5 for $\alpha=8.82$ at $\beta=811$.
The metric function $A(r)$ has a different horizon at $\ln[r]=-0.303$ in comparison to  the RN horizon at  $\ln[r]=-0.154$ and it approaches the RN metric function $f(r)$ as
$\ln[r]$ increases. Also, $\delta(r)=\ln[B(r)/A(r)]/2$ decreases as $\ln[r]$ increases, while $\delta_{\rm RN}(r)=0$ remains zero because of $B(r)/A(r)=1$ for the RN case.
From (\ref{asym-sol}), we observe a difference between $\phi(r)$ and $\phi_{\rm ml}$ for the EMS theory in the asymptotic region.
 The  other scalarized charged black holes for $\beta=258,~356,~1400,~2000$ are found similarly.

\section{Stability of scalarized charged black holes}

It turns out that the $n=0(\beta=\infty)$ black hole is stable, while the $n=1,2,\cdots(\beta\to\infty)$ black holes are unstable in the EMS theory with exponential and quadratic couplings~\cite{Myung:2019oua}.
Now, let us analyze  the stability of $n=0,1$ black holes the EMS theory with scalar mass term. For this purpose, we choose three scalar masses of  $\beta=258,~356,~811,~1400,~2000$  whose $n=0$ and $n=1$  bifurcation points are given by
$\alpha_{n=0}=\{9.619,~9.345,~8.82,~8.60,~8.493\}$ and $\alpha_{n=1}=\{56.73,~54.01,~49.15,~47.03,~45.96\}$, respectively. We focus on larger $\beta$ which provides  smaller scalar mass $m^2_\phi$ for computation.

For simplicity, we perform  radial (spherically symmetric) perturbations by choosing three perturbations
of $H_0(t,r),H_1(t,r),\delta\phi(t,r)$ as
\begin{eqnarray}
ds^2_{\rm RP}&=&-A(r)\left(1+\epsilon H_0\right)dt^2+\frac{dr^2}{B(r)\left(1
+\epsilon H_1\right)}+r^2(d\theta^2+\sin^2\theta d\chi^2),\nonumber\\
\phi&=&\phi(r)+\epsilon\delta\phi, \label{pert-metric}
\end{eqnarray}
where $A(r),~B(r),~\phi(r)$ denote a scalarized charged black hole and  $\epsilon$ is a control parameter of perturbations.
Considering the separation of variables
\begin{eqnarray}
\delta\phi(t,r)=\phi_1(r)e^{\Omega t},
\end{eqnarray}
we obtain the Schr\"{o}dinger-type equation for scalar perturbation
\begin{eqnarray}
\frac{d^2\phi_1(r)}{d r_*^2}-\Big[\Omega^2+V_{\rm SBH}(r)\Big]\phi_1(r)=0,\label{radial-pert}
\end{eqnarray}
with $r_*$ is the tortoise coordinate defined by
\begin{eqnarray}
\frac{dr_*}{dr}=\frac{1}{\sqrt{A(r)B(r)}}\label{tort}
\end{eqnarray}
and  its potential  reads as
\begin{eqnarray}
V_{\rm SBH}(r)&=&\frac{\alpha}{\beta}A\left(1+\alpha\phi^2-2\alpha^2\phi^4
+r\phi(4+5\alpha\phi^2)\phi'\right)\nonumber\\
&&-\frac{B'A}{2r}\left(-1-2\alpha+4\alpha^2\phi^2-10r\alpha\phi'\phi
+3r^2\phi'^2\right)\nonumber\\
&&-\frac{\alpha A}{r^2}(1-B)(1-2\alpha\phi^2+5r\alpha\phi\phi')+\frac{AB'(r^2\phi'^2-1)}{2r}\nonumber\\
&&+AB\phi'^2(-2-\alpha+2\alpha^2\phi^2-5\alpha r\phi\phi'+r^2\phi'^2).\label{potV}
\end{eqnarray}
\begin{figure*}[t!]
   \centering
   \includegraphics{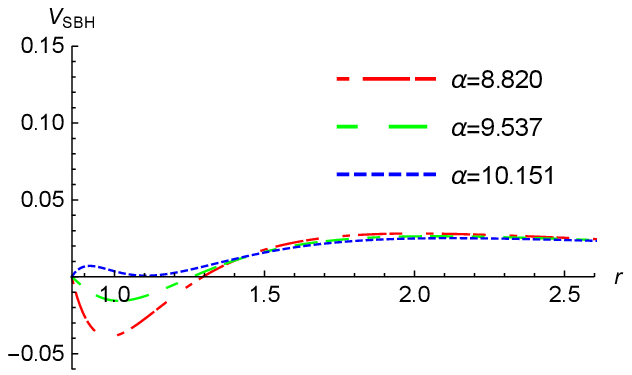}
   \hfill%
   \includegraphics{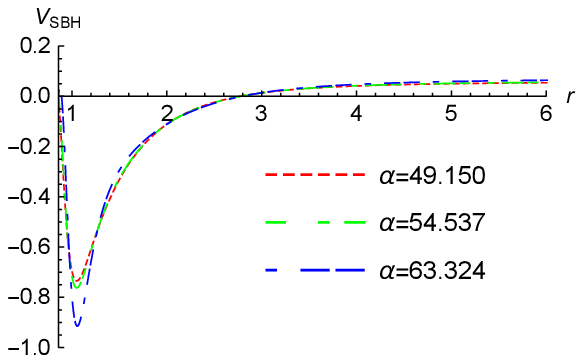}
\caption{Three scalar potentials $V_{\rm SBH}$ with $l=0$ scalar mode for $\beta=811$.   (Left) Around $n=0$ black hole. Even though they contain small negative regions outside the horizon,
these show stable black holes. (Right) Around $n=1$ black hole.
They indicate unstable black holes because their potentials  include large  negative regions outside the horizon. }
\end{figure*}
\begin{figure*}[t!]
   \centering
   \includegraphics{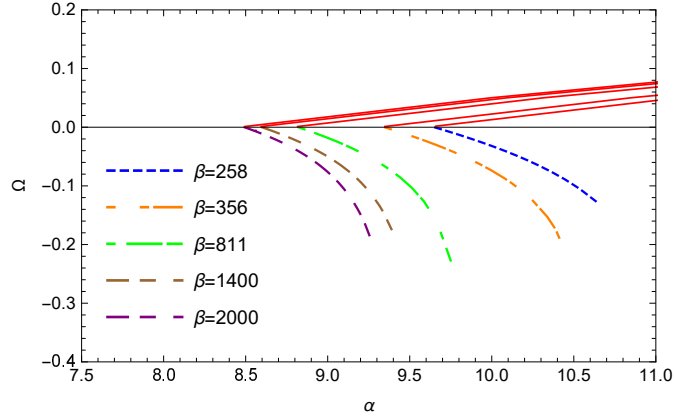}
\caption{The negative $\Omega$ is given as function of $\alpha$ for the $l=0$  scalar mode  around the $n=0$ black hole,
showing stability. Here we consider five different cases of $\beta=258$, 356, 811, 1400, and 2000. Five dotted curves start from
$\alpha_{n=0}={9.649,9.345,8.82,8.60,8.493}$. Five red lines denote the RN black holes [See Fig. 4]. }
\end{figure*}

\begin{figure*}[t!]
   \centering
   \includegraphics{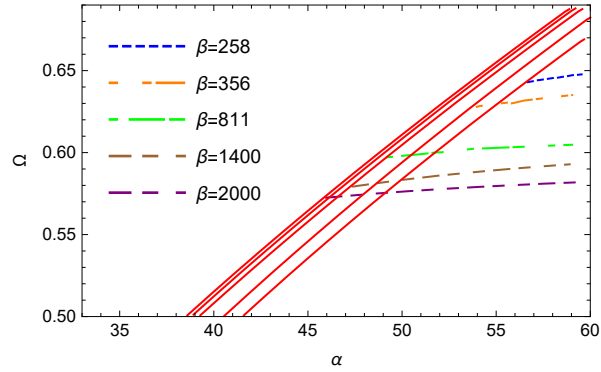}
\caption{ The positive $\Omega$ is given
as function of $\alpha$ for the $s$-mode of scalar around the $n=1$ black hole, indicating instability.
 Here we consider five different cases of $\beta=258$, 356, 811, 1400, and 2000.
Five dotted curves start from $\alpha_{n=1}={56.73,54.01,49.25,47.03,45.96}$. Five red lines represent the RN black holes. }
\end{figure*}
It is suggested  from Fig. 6 that the potentials around the $n=0$ black hole indicates small negative regions around the horizon, suggesting  the instability.
On the other hand, the potentials around the $n=1$ black hole indicates large negative regions outside  the  horizon, showing the instability.
However, the former case may be not true. The potential $V_{\rm SBH}$ with $\alpha=8.820(\beta=811)$ with  small negative region  does not imply the instability, but it might support the stability.
The linearized  scalar equation (\ref{radial-pert}) around the $n=0,1$ scalarized charged  black holes may allow  either a stable (decaying) mode with $\Omega<0$ or an unstable (growing) mode with $\Omega>0$.

We solve (\ref{radial-pert}) numerically  with imposing a boundary condition that $\phi_1(r)$ vanishes at the horizon and at infinity. We find from Figs. 7 and 8 that the $n=0$ black hole is stable against the $l=0$ scalar mode,
while  the $n=1$ black hole is unstable against the $l=0$ scalar mode.  Furthermore, we show that  that the (in) stability of $n=0(n=1)$ black holes is independent of the mass parameter $\beta$.

\section{Discussions}
One of original motivations to study this work  is to understand the difference between infinite branches in the EMS theory and  a single branch in the EW theory.
The infinite branches of $n=0,1,2,\cdots$ scalarized charged black holes in the EMS theory are not changed for $\beta> 4.4$ even for  including a scalar mass term $m^2_\phi=\alpha/\beta$, whereas these all disappear for $0<\beta \le 4.4$. This is so because the bifurcation points is determined solely  by the exponential coupling to the Maxwell term in the scalar equation (\ref{s-equa}). This implies that the role of scalar mass term provides either nothing or all bifurcation points, but it never lead  to a single branch of scalarized charged black holes.
On the other hand, the single branch is determined by the static Lichnerowicz-Ricci tensor equation [$(\vartriangle_{\rm L}+m^2_2)\delta R_{\mu\nu}=0$]  where a single bifurcation point is given by  $m_2^2=0.7677$~\cite{Lu:2017kzi}. This indicates a difference between scalar and Ricci-tensor hairs.

In this work, we have investigated the scalarized charged black holes in the EMS theory with a specific choice of scalar mass  $m^2_\phi=\alpha/\beta$.
The computing process is  as follows:  detecting instability of RN black holes$\to$ prediction of scalarized charged black holes (bifurcation points) $\to$
obtaining the $n=0,~1$ scalarized charged black holes $\to$  performing (in)stability analysis of $n=0,~1$ scalarized charged black holes.

We find  that the first two bifurcation points of  $\alpha_{n=0,1}(\beta)$ increases as $\beta$ decreases. The  RN instability is therefore harder to realize for larger scalar masses.
 We found two limits. In the massless limit of $\beta\to\infty$, $\alpha_{n=0,1}(\beta)$ approaches $\alpha_{n=0,1}=\{8.019,40.84\}$ for the EMS theory.
The other limit is given by $\alpha_{n=0}(\beta)\to \infty$, as $\beta \to 4.4$.
In other words, we have stated that unstable RN black holes exist for $\beta> 4.4$ [see the opposite bound (\ref{st-con}) for stable RN black holes].
This implies that the $n=0$  scalarized charged black hole was found for $\alpha\ge\alpha_{n=0}(\beta)$ with $8.019\le \alpha_{n=0}(\beta) <\infty $ for $\beta\in (4.4,\infty]$, showing a  shift from $\alpha_{n=0}(\beta\to\infty)=8.019$. Also,  the $n=1$  scalarized charged black hole was found for $\alpha\ge\alpha_{n=1}(\beta)$ with $40.84\le \alpha_{n=1}(\beta) <\infty $ for $\beta\in (4.4,\infty]$, showing a  shift from $\alpha_{n=1}(\beta\to\infty)=40.84$.
Interestingly, an unallowable region for scalarization is given by $0<\beta\le 4.4$ where the unstable RN black holes are never found for any $\alpha>0$.

Finally, we have shown  that the $n=0$ black hole is stable against radial perturbations, while the $n=1$ black hole is unstable.
Further, it was shown  that the stability result of $n=0,~1$ black holes is independent of the mass parameter $\beta$,
even though it changes the bifurcation points significantly.

 \vspace{1cm}

{\bf Acknowledgments}
 \vspace{1cm}

This work was supported by the National Research Foundation of Korea (NRF) grant funded by the Korea government (MOE)
 (No. NRF-2017R1A2B4002057).


\vspace{1cm}
\newpage
\begin{thebibliography}{99}
%\cite{Doneva:2017bvd}
\bibitem{Doneva:2017bvd}
  D.~D.~Doneva and S.~S.~Yazadjiev,
  %``New Gauss-Bonnet Black Holes with Curvature-Induced Scalarization in Extended Scalar-Tensor Theories,''
  Phys.\ Rev.\ Lett.\  {\bf 120}, no. 13, 131103 (2018)
  doi:10.1103/PhysRevLett.120.131103
  [arXiv:1711.01187 [gr-qc]].
  %%CITATION = doi:10.1103/PhysRevLett.120.131103;%%
  %92 citations counted in INSPIRE as of 05 Nov 2019

%\cite{Silva:2017uqg}
\bibitem{Silva:2017uqg}
  H.~O.~Silva, J.~Sakstein, L.~Gualtieri, T.~P.~Sotiriou and E.~Berti,
  %``Spontaneous scalarization of black holes and compact stars from a Gauss-Bonnet coupling,''
  Phys.\ Rev.\ Lett.\  {\bf 120}, no. 13, 131104 (2018)
  doi:10.1103/PhysRevLett.120.131104
  [arXiv:1711.02080 [gr-qc]].
  %%CITATION = doi:10.1103/PhysRevLett.120.131104;%%
  %94 citations counted in

  %\cite{Antoniou:2017acq}
\bibitem{Antoniou:2017acq}
  G.~Antoniou, A.~Bakopoulos and P.~Kanti,
  %``Evasion of No-Hair Theorems and Novel Black-Hole Solutions in Gauss-Bonnet Theories,''
  Phys.\ Rev.\ Lett.\  {\bf 120}, no. 13, 131102 (2018)
  doi:10.1103/PhysRevLett.120.131102
  [arXiv:1711.03390 [hep-th]].
  %%CITATION = doi:10.1103/PhysRevLett.120.131102;%%
  %92 citations counted in INSPIRE as of 05 Nov 2019


%\cite{Brihaye:2018grv}
\bibitem{Brihaye:2018grv}
  Y.~Brihaye and L.~Ducobu,
  %``Hairy black holes, boson stars and non-minimal coupling to curvature invariants,''
  Phys.\ Lett.\ B {\bf 795}, 135 (2019)
  doi:10.1016/j.physletb.2019.06.006
  [arXiv:1812.07438 [gr-qc]].
  %%CITATION = doi:10.1016/j.physletb.2019.06.006;%%
  %6 citations counted in INSPIRE as of 04 Jul 2019


%\cite{Macedo:2019sem}
\bibitem{Macedo:2019sem}
  C.~F.~B.~Macedo, J.~Sakstein, E.~Berti, L.~Gualtieri, H.~O.~Silva and T.~P.~Sotiriou,
  %``Self-interactions and Spontaneous Black Hole Scalarization,''
  Phys.\ Rev.\ D {\bf 99}, no. 10, 104041 (2019)
  doi:10.1103/PhysRevD.99.104041
  [arXiv:1903.06784 [gr-qc]].
  %%CITATION = doi:10.1103/PhysRevD.99.104041;%%
  %9 citations counted in INSPIRE as of 25 Jun 2019



%\cite{Doneva:2019vuh}
\bibitem{Doneva:2019vuh}
  D.~D.~Doneva, K.~V.~Staykov and S.~S.~Yazadjiev,
  %``Gauss-Bonnet black holes with a massive scalar field,''
  Phys.\ Rev.\ D {\bf 99}, no. 10, 104045 (2019)
  doi:10.1103/PhysRevD.99.104045
  [arXiv:1903.08119 [gr-qc]].
  %%CITATION = doi:10.1103/PhysRevD.99.104045;%%
  %3 citations counted in INSPIRE as of 25 Jun 2019

%\cite{Herdeiro:2018wub}
\bibitem{Herdeiro:2018wub}
  C.~A.~R.~Herdeiro, E.~Radu, N.~Sanchis-Gual and J.~A.~Font,
  %``Spontaneous Scalarization of Charged Black Holes,''
  Phys.\ Rev.\ Lett.\  {\bf 121}, no. 10, 101102 (2018)
  doi:10.1103/PhysRevLett.121.101102
  [arXiv:1806.05190 [gr-qc]].
  %%CITATION = doi:10.1103/PhysRevLett.121.101102;%%
  %28 citations counted in INSPIRE as of 04 Jul 2019

%\cite{Fernandes:2019rez}
\bibitem{Fernandes:2019rez}
  P.~G.~S.~Fernandes, C.~A.~R.~Herdeiro, A.~M.~Pombo, E.~Radu and N.~Sanchis-Gual,
  %``Spontaneous Scalarisation of Charged Black Holes: Coupling Dependence and Dynamical Features,''
  Class.\ Quant.\ Grav.\  {\bf 36}, no. 13, 134002 (2019)
  doi:10.1088/1361-6382/ab23a1
  [arXiv:1902.05079 [gr-qc]].
  %%CITATION = doi:10.1088/1361-6382/ab23a1;%%
  %5 citations counted in INSPIRE as of 04 Jul 2019


%\cite{Myung:2019oua}
\bibitem{Myung:2019oua}
  Y.~S.~Myung and D.~C.~Zou,
  %``Stability of scalarized charged black holes in the Einstein–Maxwell–Scalar theory,''
  Eur.\ Phys.\ J.\ C {\bf 79}, no. 8, 641 (2019)
  doi:10.1140/epjc/s10052-019-7176-7
  [arXiv:1904.09864 [gr-qc]].
  %%CITATION = doi:10.1140/epjc/s10052-019-7176-7;%%
  %3 citations counted in INSPIRE as of 12 Aug 2019



%\cite{Lu:2015cqa}
\bibitem{Lu:2015cqa}
  H.~Lu, A.~Perkins, C.~N.~Pope and K.~S.~Stelle,
  %``Black Holes in Higher-Derivative Gravity,''
  Phys.\ Rev.\ Lett.\  {\bf 114}, no. 17, 171601 (2015)
  doi:10.1103/PhysRevLett.114.171601
  [arXiv:1502.01028 [hep-th]].
  %%CITATION = doi:10.1103/PhysRevLett.114.171601;%%
  %94 citations counted in INSPIRE as of 01 Jul 2019


%\cite{Lu:2017kzi}
\bibitem{Lu:2017kzi}
  H.~Lü, A.~Perkins, C.~N.~Pope and K.~S.~Stelle,
  %``Lichnerowicz Modes and Black Hole Families in Ricci Quadratic Gravity,''
  Phys.\ Rev.\ D {\bf 96}, no. 4, 046006 (2017)
  doi:10.1103/PhysRevD.96.046006
  [arXiv:1704.05493 [hep-th]].
  %%CITATION = doi:10.1103/PhysRevD.96.046006;%%
  %17 citations counted in INSPIRE as of 01 Jul 2019


%\cite{Myung:2018vug}
\bibitem{Myung:2018vug}
  Y.~S.~Myung and D.~C.~Zou,
  %``Instability of Reissner–Nordström black hole in Einstein-Maxwell-scalar theory,''
  Eur.\ Phys.\ J.\ C {\bf 79}, no. 3, 273 (2019)
  doi:10.1140/epjc/s10052-019-6792-6
  [arXiv:1808.02609 [gr-qc]].
  %%CITATION = doi:10.1140/epjc/s10052-019-6792-6;%%
  %11 citations counted in INSPIRE as of 01 Jul 2019



\bibitem{Zerilli:1974ai}
  F.~J.~Zerilli,
  %``Perturbation analysis for gravitational and electromagnetic radiation in a reissner-nordstroem geometry,''
  Phys.\ Rev.\ D {\bf 9}, 860 (1974).
  doi:10.1103/PhysRevD.9.860
  %%CITATION = doi:10.1103/PhysRevD.9.860;%%
  %127 citations counted in INSPIRE as of 16 Jul 2018

%\cite{Moncrief:1974gw}
\bibitem{Moncrief:1974gw}
  V.~Moncrief,
  %``Odd-parity stability of a Reissner-Nordstrom black hole,''
  Phys.\ Rev.\ D {\bf 9}, 2707 (1974).
  doi:10.1103/PhysRevD.9.2707
  %%CITATION = doi:10.1103/PhysRevD.9.2707;%%
  %119 citations counted in INSPIRE as of 16 Jul 2018

%\cite{Moncrief:1974ng}
\bibitem{Moncrief:1974ng}
  V.~Moncrief,
  %``Stability of Reissner-Nordstrom black holes,''
  Phys.\ Rev.\ D {\bf 10}, 1057 (1974).
  doi:10.1103/PhysRevD.10.1057
  %%CITATION = doi:10.1103/PhysRevD.10.1057;%%
  %95 citations counted in INSPIRE as of 16 Jul 2018

%\cite{Moncrief:1975sb}
\bibitem{Moncrief:1975sb}
  V.~Moncrief,
  %``Gauge-invariant perturbations of Reissner-Nordstrom black holes,''
  Phys.\ Rev.\ D {\bf 12}, 1526 (1975).
  doi:10.1103/PhysRevD.12.1526
  %%CITATION = doi:10.1103/PhysRevD.12.1526;%%
  %83 citations counted in INSPIRE as of 16 Jul 2018




\end{thebibliography}
\end{document}